# Modelling of Si-B-N ceramics

A.Hannemann[1], J.C.Schön[1], C.Oligschleger[2], M.Jansen[1]
[1] Max-Planck-Institut für Festkörperforschung, Heisenbergstr. 1, 70569 Stuttgart
[2] Institut für Theoretische Chemie Universität Bonn, Wegelerstr. 10, 53115 Bonn

Abstract

We present results of computer simulations of amorphous $Si_3B_3N_7$ ceramics. The computed pair correlation functions are in satisfactory agreement with the experimental results. Regarding the structural properties of the B-N, Si-N and N-(Si/B) coordination spheres, our results are in full agreement with the experimental data: We confirm that silicon is tetrahedrally coordinated by nitrogen, and boron is trigonally-planar coordinated by nitrogen. Furthermore we find that the coordination of nitrogen by the cations is also trigonal-planar. The simulation of the relaxations and annealing processes of this new substance indicate the existence of rather stable $BN_3$ building units, while the $SiN_4$ building units are comparatively malleable.

# 1 Introduction

Glasses and ceramics belong to the oldest manufactured materials. Despite long experience in the production of glasses (typically by a heat and quench procedure) there is still a lack of information about the properties of these "classic" materials. Furthermore, recent experimental research on high-quality ceramics has shown that these new materials are only available via precursor routes with subsequent polymerization of these precursors [1]. Since the widespread technical relevance of such materials has become apparent, the demand for an understanding of structural and dynamical properties of both new and old amorphous materials has dramatically increased. But due to the lack of periodicity, such information can only be gained using a combination of many sophisticated experimental (e.g. X-ray/neutron scattering, NMR/ESR-spectroscopy) and simulation techniques.
Thus we have performed Monte-Carlo simulations [2] of amorphous $Si_3B_3N_7$ ceramics using different configurations as the starting points for our simulations. The next section describes the computational aspects of the simulations, structural properties are analyzed in section three, and some aspects of the relaxation dynamics are described in section four.

# 2 Computational Aspects

## 2.1 Starting configurations

In order to gain insight into the structural properties of these amorphous Si-B-N ceramics by computer simulations we used the following configurations as starting points for our investigations:
a) Melts of hypothetical crystalline $Si_3B_3N_7$ polymorphs [3] generated by molecular-dynamics simulations ("Melts"), configurations generated by a random-closed packing scheme [4] ("RCP"), and a configuration generated by an arrangement of crystalline fragments of the constituting binary phases, hexagonal-BN and $\beta$-$Si_3N_4$ ("CF"). The "Melt" and the "RCP" starting configurations consisted of 702 atoms, whereas the configuration obtained from an arrangement of crystalline fragments with a diameter of about 5Å consisted of 1144 atoms. We employed periodic boundary conditions, with side lengths of the cubic simulation cells $\approx 15 - 20$Å.

## 2.2 Computational details

### 2.2.1 General Monte-Carlo techniques

During the course of Monte-Carlo type-simulations (*Relaxations / Simulated Annealing*) one changes the current configuration randomly to a trial configuration. A trial configuration is accepted with a probability proportional to the Boltzmann factor $\exp(-\Delta E/kT)$, where T is a (fictive) temperature. In the case of relaxations, the (fictive) temperature is kept constant for the whole observation time $\tau_{obs}$, and the system evolves at the given temperature, whereas in *Simulated Annealing* [5] one starts with an initial temperature $T_0$ and repeatedly lowers the fictive temperature after a simulation time $\tau_s$ according to some (usually) predefined function T(t) until the predefined final temperature $T_f$ (typically $T_f \approx 0$) is reached.
In this work we either change the atomic position of a randomly chosen atom by a random vector $\vec{r}_{max}$ of maximum length $d_{max}$ (99% of the time) or we change the cell volume (1% of the time). The maximum length $d_{max}$ was varied during the simulations in order to achieve an acceptance rate of about 50%. Either configuration change or attempt is called a MC-move. A sequence of N MC-moves (N is the number of atoms) constitutes a MC-cycle. The allowed MC-moves taken together define the moveclass of the MC-simulations.

### 2.2.2 Parameters of the simulations

The configurations obtained from the melt and the RCP-scheme were allowed to relax at constant temperatures $T_0$ =1000K, 2000K, 3000K for 5000 MC-cycles. The interactions between the atoms were described by a potential (A) given by Gastreich et al. [6]. After this "pre-equilibration procedure", the temperature was lowered by 30K every 1000 MC-cycles (linear cooling schedule) until $T_f$ = 0K was reached. The final configurations of these simulated annealing runs were further optimized by conjugate gradients using the program GULP [7]. Here a second (more refined) interaction potential (B) given by Gastreich et. al. [8] was employed.
The configuration obtained from the crystalline fragments was first relaxed at 1K for 5000 MC-cycles. Subsequently the temperature was raised by 30K every 1000 MC-cycles (linear heating schedule), up to a final temperature of about 210K. The configurations obtained were again optimized by conjugate gradients, applying the second interaction potential (B).

## 3 Structure of silicon-boron-nitride networks

We characterize the structure of the Si-B-N networks by their total pair correlation functions G(r) and their partial pair distribution functions $g_{\alpha\beta}(r)$.
G(r) is obtained experimentally from the Fourier transform of the measured structure factor S(Q). We calculate the total pair-correlation function G(r) from the simulated structures using the well-known Debye formula [9], which connects intensity data to interatomic distances. The pair distribution functions $g_{\alpha\beta}(r)$ from the simulated structures are calculated using the following definition: $n_{\alpha\beta}\Delta r = 4\pi\rho_N c_\beta g_{\alpha\beta}(r)$, where $n_{\alpha\beta}\Delta r$ is the number of particles of element $\beta$ in a spherical shell with thickness $\Delta r$ and radius r around a particle of element $\alpha$, averaged over all atoms $\alpha$. $\rho_N = N/V$ is the number density and $c_\beta = N_\beta/N$ is the concentration of element $\beta$. Note that with these definitions $g_{\alpha\beta}(r)$ resembles a normalized counting of distances, whereas the total pair correlation function is a weighted (by the atomic scattering factors) counting of interatomic distances. Nevertheless peak locations are the same in both functions.

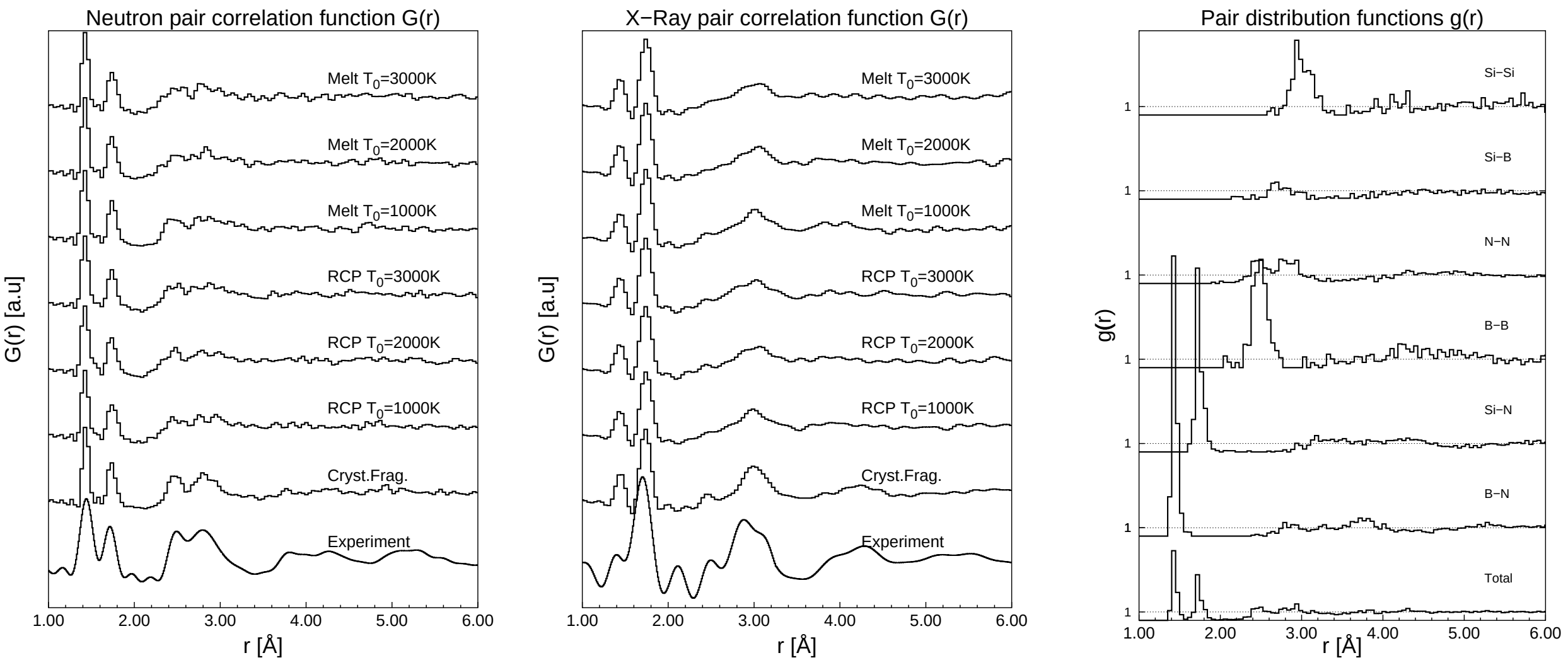

Figure 1: **Left**: Pair correlation function G(r) from neutron scattering. **Middle**: Pair correlation function G(r) from x-ray scattering. "Experiment" denotes the experimental curves, "Cryst.Frag." is obtained from the crystal fragment starting configuration at 200K, "RCP" is obtained from the random close packing scheme and "Melt" corresponds to the configuration that resulted from the MD-melt. "Pre-equilibration" temperatures are denoted by $T_0$. **Right**: Partial pair-distribution-functions for model generated from crystalline fragments which showed the best agreement with the experimental data (see figures to the left).

After inspection of the pair distribution functions, we defined the first coordination sphere of silicon and boron as the set of atoms whose distance to the cebteral atoms boron and silicon lie between the cutoff-radii: $\mathrm{r_{min}(B)} = 1.2$Å, $\mathrm{r_{max}(B)} = 1.7$Å, and $\mathrm{r_{min}(Si)} = 1.4$Å, $\mathrm{r_{max}(Si)} = 2.1$Å, respectively. From these coordination spheres we derive the angular distribution functions for silicon and boron. Nitrogen coordination spheres are also calculated, with the cutoff-radii $\mathrm{r_{min}(N)} = 1.2$Å and $\mathrm{r_{max}(N)} = 2.1$Å. In order to study the cation-cation coordinations we have also calculated the appropriate coordination numbers in the following manner: For each cation we investigate its nearest neighbors (bridging atoms) and identify the nearest neighbors of the bridging atoms as the coordinating atoms of the cation (omitting the original cation itself, of course).

## 3.1 Results of the simulation

In figure 1 we show typical neutron and X-ray pair correlation functions in comparison to the experimental data [10] in the range from 1.0 to 6.0Å. One clearly notices four distinct peaks at approximately 1.43, 1.75, 2.60 and 2.90Å in the pair correlation function obtained from the Fourier transform of the total structure factor S(Q) (for details c.f. tables 1, 2, 3). The origin of the first two peaks can clearly be attributed to B-N and Si-N distances, whereas the origin of subsequent peaks may stem from N-N, B-B, Si-B and Si-Si distances. The partial pair distribution functions (c.f. figure 1) disclose the origin of these peaks for the configuration generated from the arrangement of crystalline fragments. Note the good agreement in the total pair correlation functions of the simulation results with the experimental curves, in particular for the configuration based on the arrangement of crystalline fragments. The peak at about 2.60Å can be attributed to both N-N and B-B interatomic distances and the peak at about 2.90Å to Si-Si interatomic distances, respectively. Furthermore, Si-B and a second N-N interatomic distance contribute to the second double-peak.

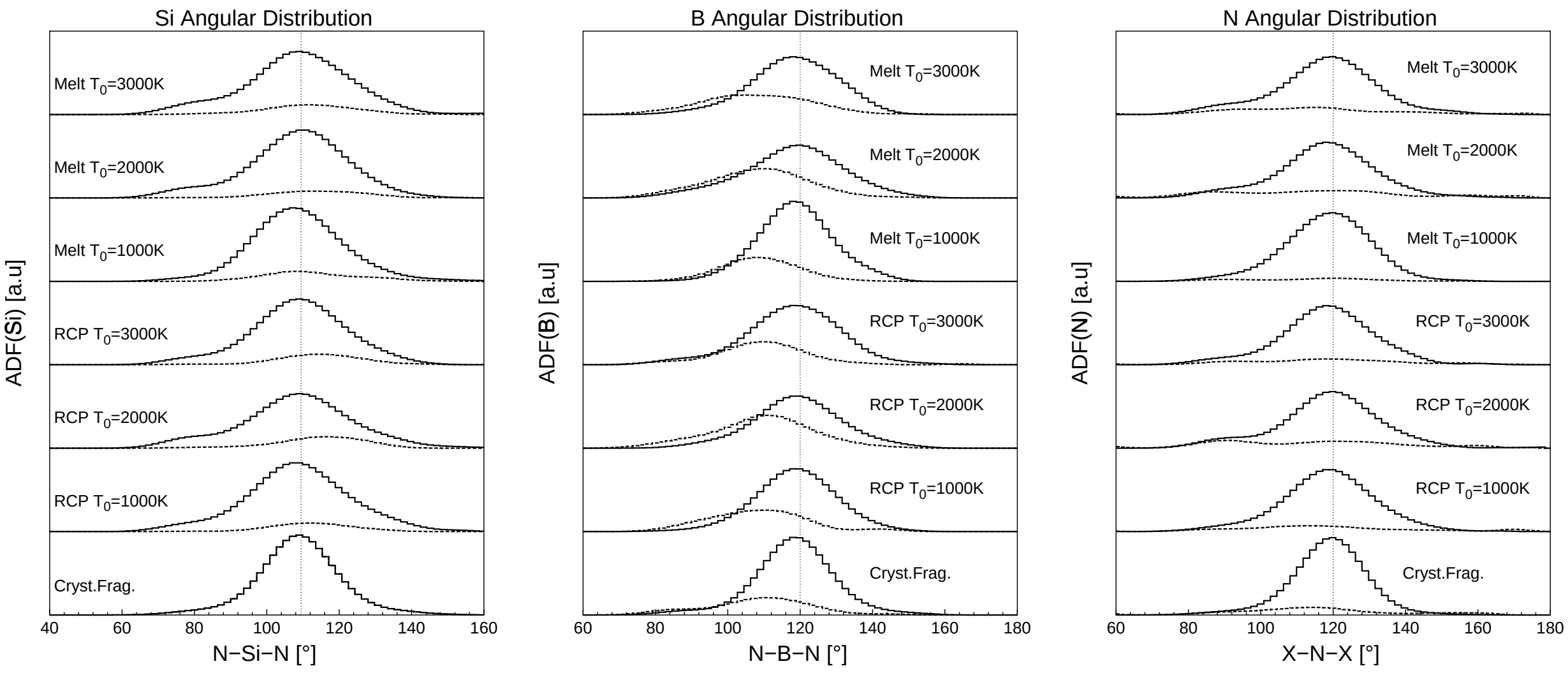

Figure 2: Angular distribution functions ADF for Si (Left), B (Middle) and N (Right). Upper curves in each plot corresponds to $SiN_4$, $BN_3$ and $NX_3$ (X=Si/B), respectively. Lower curves are the angular distribution functions for (the rarely encountered) $SiN_3$, $BN_4$ and $NX_4$ coordination spheres, respectively. Vertical dotted lines indicate angles of 109.47° and 120°, respectively.

Experimental data for the first coordination spheres of boron and silicon suggested three-fold and four-fold coordination (MCN(B) = 2.95, MCN(Si) = 3.8)[10] [1]. The nitrogen coordination number in the $Si_3B_3N_7$ ceramic has recently been recorded to be three [11], but the distribution of $NSi_{3-y}B_y$ (y = 0,1,2,3) has not yet been determined experimentally. Therefore we have analyzed the coordinations spheres of all atomic species and calculated their angular distribution functions (ADF) (c.f. figure 3.1 and tables 1, 2, 3). The mean coordination numbers of boron and silicon are in agreement with experimental data as are their mean bond lengths (MBL) (Experiment: MBL(B-N) = 1.44Å, MBL(Si-

[1]MCN(X) is the mean coordination number of element X

| Start | $T_0$[K] | $MCN_B$ | $MBL_B$ | $BN_2$ | $BN_3$ | $BN_4$ |
|---|---|---|---|---|---|---|
| RCP | 3000 | 3.12 | 1.43 | 1.85 | 83.95 | 14.19 |
| RCP | 2000 | 3.22 | 1.44 | 0.61 | 75.92 | 23.45 |
| RCP | 1000 | 3.14 | 1.43 | 1.85 | 81.48 | 16.66 |
| MELT | 3000 | 3.15 | 1.43 | 1.23 | 82.09 | 16.66 |
| MELT | 2000 | 3.18 | 1.44 | 1.23 | 79.01 | 19.75 |
| MELT | 1000 | 3.13 | 1.43 | 0.00 | 87.03 | 12.96 |
| CF | 200 | 3.12 | 1.44 | 0.37 | 87.12 | 12.50 |

Table 1: Mean coordination number $MCN_B$, mean bond length $MBL_B$ and occurrence of coordination spheres $BN_2$, $BN_3$ and $BN_4$ (in percent). The column "Start" corresponds to the different starting configuration mentioned in section 2, "pre-equilibration" temperatures $T_0$ are given in the second column.

| Start | $T_0$ [K] | $MCN_{Si}$ | $MBL_{Si}$ | $SiN_3$ | $SiN_4$ | $SiN_5$ |
|---|---|---|---|---|---|---|
| RCP | 3000 | 3.76 | 1.75 | 25.30 | 72.84 | 1.23 |
| RCP | 2000 | 3.75 | 1.75 | 25.30 | 71.60 | 1.85 |
| RCP | 1000 | 3.83 | 1.75 | 17.90 | 80.86 | 0.61 |
| MELT | 3000 | 3.79 | 1.75 | 21.60 | 77.16 | 1.23 |
| MELT | 2000 | 3.85 | 1.75 | 17.28 | 80.24 | 1.85 |
| MELT | 1000 | 3.82 | 1.75 | 18.51 | 80.24 | 0.61 |
| CF | 200 | 3.80 | 1.74 | 20.45 | 78.78 | 0.75 |

Table 2: Mean coordination number $MCN_{Si}$, mean bond length Si-N $MBL_{Si}$ and occurrence of coordination spheres $SiN_3$, $SiN_4$ and $SiN_5$ (in percent, differences to 100% are caused by $SiN_2$). The column "Start" corresponds to the different starting configuration mentioned in section 2, "pre-equilibration" temperatures $T_0$ are given in the second column.

| Start | $T_0$[K] | $MCN_N$ | $NSi_3$ | $NSi_2B_1$ | $NSi_1B_2$ | $NB_3$ |
|---|---|---|---|---|---|---|
| RCP | 3000 | 3.02 | 10.58 | 38.09 | 30.42 | 2.64 |
| RCP | 2000 | 3.08 | 9.78 | 35.45 | 29.89 | 3.43 |
| RCP | 1000 | 3.04 | 10.05 | 41.27 | 28.04 | 5.02 |
| MELT | 3000 | 3.07 | 9.25 | 40.74 | 25.39 | 3.43 |
| MELT | 2000 | 3.13 | 11.90 | 33.33 | 28.04 | 2.91 |
| MELT | 1000 | 3.00 | 11.37 | 40.47 | 34.12 | 3.70 |
| CF | 200 | 3.03 | 34.25 | 12.98 | 12.82 | 21.59 |

Table 3: Mean coordination $MCN_N$ and occurrence of coordination spheres $NSi_3$, $NSi_2B$, $NSi_1B_2$ and $NB_3$ (in percent, differences to 100% are caused by $NX_2$ or $NX_4$ coordinations). The column "Start" corresponds to the different starting configuration mentioned in section 2, "pre-equilibration" temperatures $T_0$ are given in the second column.

| Start | $T_0$ [K] | B-B | B-Si | Si-B | Si-Si |
|---|---|---|---|---|---|
| RCP | 1000 | 2.61 | 3.85 | 3.87 | 3.85 |
| RCP | 2000 | 2.72 | 3.95 | 3.96 | 3.61 |
| RCP | 3000 | 2.58 | 3.74 | 3.74 | 3.65 |
| MELT | 2000 | 2.79 | 3.92 | 3.94 | 3.88 |
| MELT | 3000 | 2.56 | 3.95 | 3.96 | 3.64 |
| MELT | 1000 | 2.44 | 3.87 | 3.87 | 3.76 |
| CF | 200 | 4.93 | 1.57 | 1.58 | 5.84 |

Table 4: Cation-Cation coordination (B-B, B-Si, Si-B and Si-Si). Column "Start" corresponds to the different starting configurations mentionend in section 2, "pre-equilibration" temperatures $T_0$ are given in the second column.

N) = 1.73Å). Our simulations show that nitrogen is mainly three-fold coordinated. The geometry of these coordination spheres can be deduced from the corresponding angular distribution functions of the coordination spheres $SiN_4$, $BN_3$ and $NX_3$, $(X = Si/B)$, respectively (c.f. figure 3.1). The angular distribution function of silicon shows a maximum at 109.5°, thus indicating the tetrahedral geometry. For boron and nitrogen one observes a maximum at 120.0°. In connection with the mean coordination number of about three for both atomic species, this suggests a three-fold trigonal-planar coordination for boron and nitrogen.

The distribution of nitrogen coordination spheres strongly depends on the starting configurations of the simulations: Even though the mean coordination numbers are about three for all structural models, the fraction of pure $NX_3$ (X either Si or B) building units in comparison to mixed $NSi_{3-y}B_y$ (y = 1,2) building units changes from 10% ("Melt" and "RCP") to about 35% for the models obtained from the crystal fragment arrangements. A similar behaviour is found in the cation-cation coordination numbers (c.f. table 4). The configurations obtained from the "RCP" and the "Melt" starting configurations show no preference for like-like cation-cation coordination (e.g. Si-Si vs. Si-B), whereas the models obtained from the "CF" starting configuration do indeed show a marked preference (MCN(Si-Si) = 5.84 vs. MCN(Si-B) = 1.58).

# 4 Dynamic properties

Since there exists little experimental knowledge about the reasons for the stability of the networks at elevated temperatures, we have performed Monte-Carlo relaxations at different constant temperatures using the arrangement of crystalline fragments as the starting configuration. In addition we have gathered some information about those parts of configuration space that are accessible from the starting configuration.

The relaxations were performed at temperatures of 500K, 750K, 1000K, 1250K, 1750K

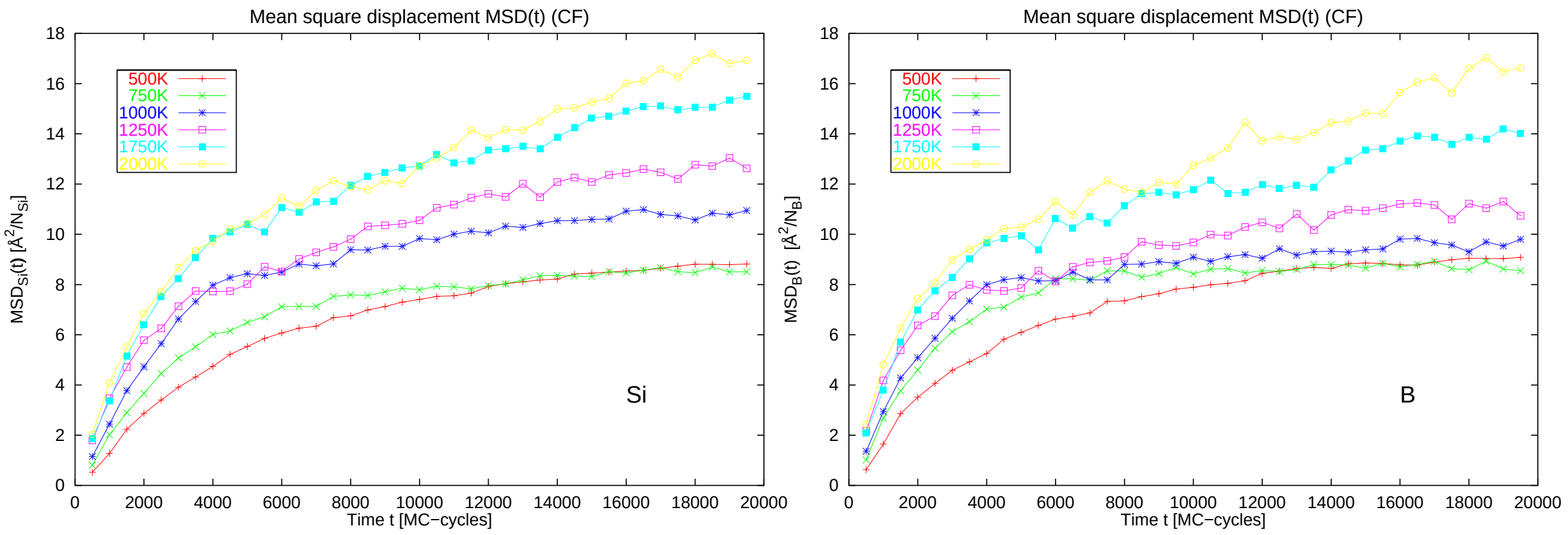

Figure 3: Mean square displacements for silicon (Left) and boron (Right) at the temperatures given in the legend.

and 2000K. For each temperature, averages were taken over three independent Monte-Carlo runs. The length of each run was $2 \times 10^4$ MC-cycles applying the same moveclass as in the simulated annealing runs mentioned in section two.

In order to study the exploration of configuration space, we use the mean square displacement MSD(t) with the definition $MSD(t) = \sum_i^N (\vec{r_i}(t) - \vec{r_i}(0))^2$. We note that the element specific MSD yields information about the mobility of the different elements. We monitor the structural stability of the first two coordination spheres (B-N and Si-N) by the time evolution of the mean coordination number of boron and silicon. In addition we consider the "survival" probability of B-N- and Si-N-bonds at time $t$ with respect to their initial occurence at time $t_0$, $P_B(t, t_0)$ = (No. of bonds at time t that were present at time $t_0$) / (No. of bonds at time $t_0$). An important feature of the Si-B-N ceramics appears to be their synthesis via precursors consisting of Si-N-B fragments [1]. We have therefore analyzed the networks with the goal of determining the number of Si-N-B fragments that build up the network in order to check to what extent our structures might be consistent with an "Aufbauprinzip" of Si-N-B-fragments [12]. Figure 4 shows the mean square displacements of silicon (Left) and boron (Right). Each graph shows typical features: after a short equilibration time of about 3000 MC-cycles accommodated by a sharp increase in the MSD(t)-curves, the curves start to grow rather slowly. From the temperature dependence of the MSD-curves, one clearly notices that the average distance travelled by each cation becomes larger as the temperature rises. In addition the synopsis of both graphs reveals that silicon atoms move further away from their position at t=0 than do the boron atoms. The "survival" probability of Si-N and B-N bonds is depicted in figure 4. One clearly recognizes a temperature dependence of the "survival" probability. As expected, the higher the temperature the lower the probability of survival becomes. The higher stability of the B-N bonds becomes obvious, if one considers both figures for temperatures $T \leq 1250K$. The probability for an Si-N bond to survive, is $\approx 0.6$, while the B-N-bonds survive with a probability of $\approx 0.85$. For higher temperatures the corresponding values are $\approx 0.25$ and $\approx 0.5$, respectively. We therefore conclude that the B-N bonds are more stable than the Si-N bonds in these networks. This conclusion is confirmed by monitoring the mean-coordination numbers for Si-N and B-N. The left graph of figure 5 shows the time dependence of the mean coordination numbers for silicon and boron (depicted as difference to ideal coordination numbers $CN_{id}$: $CN_{id}(Si) = 4, CN_{id}(B) = 3$). From this figure we also conclude that the B-N-bonds are much stronger than the Si-N bonds, in agreement with the order of decomposition temperatures of crystalline BN and $Si_3N_4$, $T_d(Si_3N_4) \approx 1800K$, $T_d(BN) \approx 2000K$, respectively. In addition we find that the mean coordination number for silicon does not change dramatically with time and temperature on the time scale of these simulations, whereas the network tends to "heal" via the generation of B-N-bonds. As soon as these bonds have emerged, the silicon atoms remain in their non-optimal arrangements. This is chemically reasonable, since the kinetics of B-N bond formation during the aminolysis stage is expected to be faster than the Si-N bond formation.

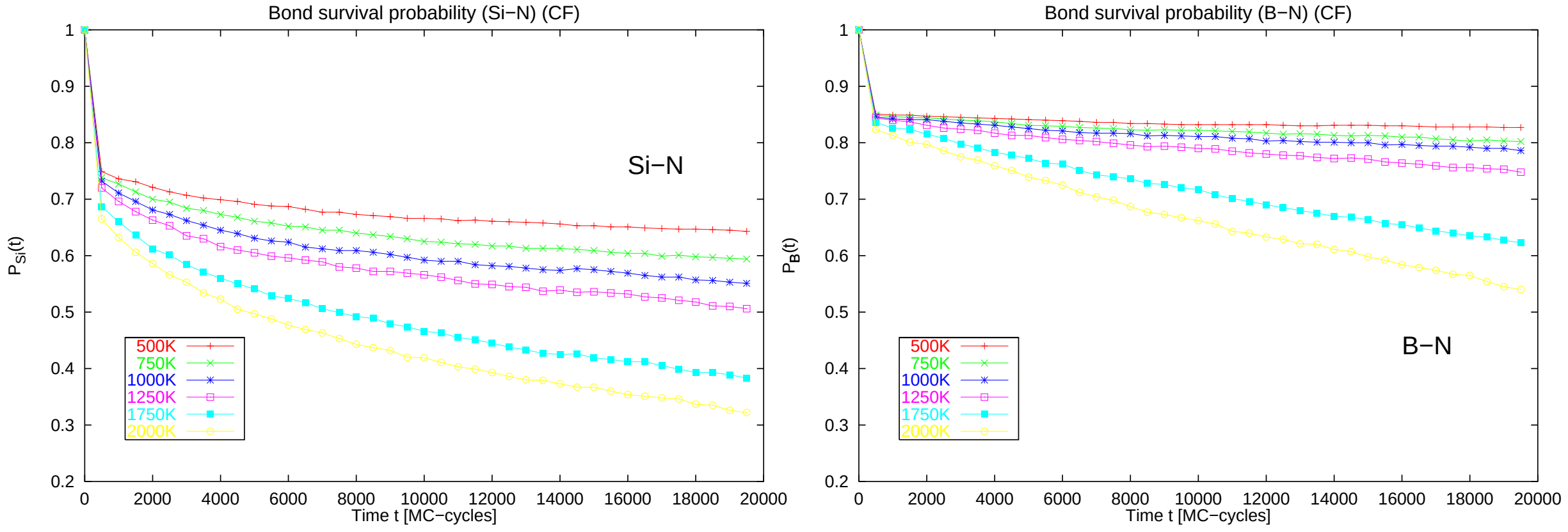

Figure 4: Bond survival probability for silicon (Left) and boron (Right) at the temperatures given in the legend.

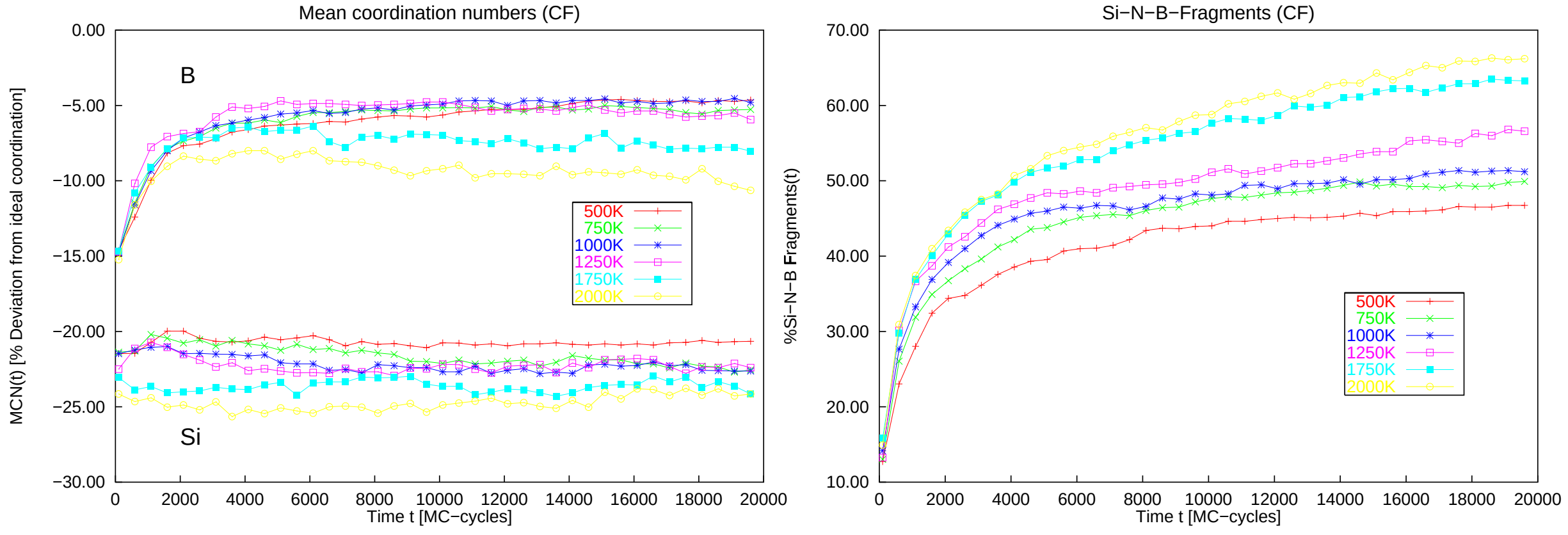

Figure 5: Percentage of existing Si-N-B-fragments (Right) and mean coordination numbers MCN for silicon and boron (Left) at the temperatures given in the legend.

As mentioned above, Si-N-B precursor fragments seem to play in important role in the stability of these networks. Thus we have also monitored the number of Si-N-B-fragments that build up the network (c.f. figure 5 Right). We observed an increase in the number of these fragmens with time. In connection with a rapid drop in energy associated with the formation of such units (not shown) we can conclude that these fragments should play an important role in the stability of the networks.

# 5 Summary

In this study we confirmed the experimental results concerning the first coordination spheres of amorphous $Si_3B_3N_7$ networks by means of Monte-Carlo simulations. Basic building units are trigonal-planar $BN_3$-units and tetrahedral $SiN_4$-units. In addition to this we showed that nitrogen is trigonally-planar coordinated. The comparison of experimental and simulated total pair-correlation functions showed that those models obtained from the arrangement of crystal fragments and subsequent relaxation constitute the best structural models of $Si_3B_3N_7$ so far. One characteristic structural feature of these models is the existence of a high percentage of like-like cation-cation coordination.

Our results further suggest that the stability of amorphous $Si_3B_3N_7$ networks may be explained by the existence of stable $BN_3$ building units in comparison to the weaker $SiN_4$ building units and strong *Si-N-B* linking units. From our simulations we also conclude that the $SiN_4$ units serve as that part of the network that is capable of reacting to some kind of external stress, whereas the $BN_3$-units serve as rigid "network-builders" and are generated rather quickly during the synthesis and relaxation.

# 6 Acknowledgement

Financial support by the Deutsche Forschungsgemeinschaft (Sonderforschungsbereich 408) is gratefully acknowledged.